\documentstyle[epsf]{article}
\begin{document}
\title{Entangled coherent states: teleportation and decoherence}
\author{S.J. van Enk$^1$ and O. Hirota$^2$\\
$^1$Bell Laboratories,
Room 2C-401\\
600-700 Mountain Ave\\
Murray Hill NJ 07974\\
$^2$Research Center for Quantum Communications\\
Tamagawa University, Tokyo, Japan}
\maketitle
\abstract{When a superposition $(|\alpha\rangle-|-\alpha\rangle)$ of two coherent states with opposite phase falls upon a 50-50 beamsplitter, the resulting state is entangled. Remarkably, the amount of entanglement is exactly 1 ebit, irrespective of $\alpha$, as was recently discovered by O. Hirota and M. Sasaki. Here we discuss decoherence properties of such states and give a simple protocol that teleports one qubit encoded in Schr\"odinger cat states.}
\section{Introduction}
Entangled states are useful for quantum information processing, but are hard to produce and tend to decohere fast. For quantum communication purposes entangled states of the electromagnetic (em) field are of particular interest. 
Such states can be used, e.g., for quantum key distribution \cite{qkd} and teleportation \cite{telep}.

There are different types of entanglement for light fields. For instance, in the teleportation
experiments from the Innsbruck group \cite{dik}, it is the polarization directions of single photons that are entangled. An example of a polarization-entangled state is
\begin{equation}\label{pol}
|\psi\rangle_{1,2}=(|\updownarrow\rangle_1|\leftrightarrow\rangle_2-|\leftrightarrow\rangle_1 |\updownarrow\rangle_2)/\sqrt{2},
\end{equation}
where $|\leftrightarrow\rangle$ and $|\updownarrow\rangle$ denote single-photon states with two orthogonal polarization directions and the subscripts 1 and 2 refer to different (spatial) modes of the em field.
The state (\ref{pol}) has one ebit of entanglement \cite{ent} and consequently can be used to teleport one qubit encoded in polarization.

 In the Caltech teleportation experiment \cite{akira},  two em field modes are entangled with respect to photon numbers and the state used for teleportation is a two-mode squeezed state \cite{qo} 
of the form 
\begin{equation}\label{r}
|\psi_r\rangle=\frac{1}{\cosh r}\sum_{n=0}^\infty (\tanh r)^n |n\rangle |n\rangle,
\end{equation}
where $|n\rangle$ denotes a state with $n$ photons and $r$ is a real parameter (giving the amount of squeezing in the fields used to produce $|\psi_r\rangle$). The state (\ref{r}) possesses
an amount  of entanglement equal to \cite{enk}
\begin{equation}
E(r)=\cosh^2 r \log_2 (\cosh^2 r)-\sinh^2 r\log_2 (\sinh^2 r),
\end{equation}
which can lie anywhere between zero (for $r\rightarrow 0$)  and infinity (for $r\rightarrow\infty$). In principle, $E(r)$ qubits encoded in a state of the form
$\sum_n \alpha_n |n\rangle$ can be teleported with $|\psi_r\rangle$.

In this paper we discuss a third type of entangled states of two modes of the electromagnetic field. 
They are parametrized by a complex parameter $\alpha$,
\begin{equation}\label{start}
|H_{\alpha}\rangle_{1,2}=(|\alpha\rangle_1|\alpha\rangle_2-|-\alpha\rangle_1|-\alpha\rangle_2)
/\sqrt{N_{\alpha}},
\end{equation}
where $|\alpha\rangle$ is a coherent state.
Entanglement properties of states of this form were considered in \cite{hirota1,hirota2}, although other properties, such as nonclassical photon statistics and squeezing properties were considered before \cite{chai}. Similar states but with a different relative phase factor, which does make a crucial difference (see below and \cite{knight}) were considered in \cite{yeazell,filip} and earlier in \cite{sanders}. A related family of states, including $|H_{\alpha}\rangle$,  was discussed in \cite{munro} in the context of ion traps. In spite of the crucial difference the phase factor makes, we denote the entangled states (\ref{start}) as ``entangled coherent states'', the term used in \cite{sanders}. 
The normalization factor $N_{\alpha}$ is
\begin{equation}
N_{\alpha}=2-2\exp(-4|\alpha|^2),
\end{equation}
and for later use we abbreviate
\begin{equation}
c_\alpha\equiv\langle \alpha|-\alpha\rangle=\exp(-2|\alpha|^2).
\end{equation}
The state (\ref{start}) can be expanded in the photon-number state basis as
\begin{equation}
|H_\alpha\rangle_{1,2}=\frac{2\exp(-|\alpha|^2)}{\sqrt{N_\alpha}}\sum_{n,m|n+m \,{\rm odd}} \frac{\alpha^{n+m}}{\sqrt{n!m!}}|n\rangle_1|m\rangle_2.
\end{equation}
Note the difference between $|H_\alpha\rangle$ and the two-mode squeezed state $|\psi_r\rangle$: in particular, in the former the total number of photons is always odd, whereas in the latter it is always even. We also note that the average number of photons in the state $|H_\alpha\rangle$ is
\begin{equation}
\langle N\rangle=2|\alpha|^2\frac{1+c_\alpha^2}{1-c_\alpha^2},
\end{equation}
which reduces to 1 in the limit $\alpha\rightarrow 0$ and to $2|\alpha|^2$ for $|\alpha|\rightarrow\infty$.
The state (\ref{start}) can be produced from a Schr\"odinger cat state
\begin{equation}
(|\sqrt{2}\alpha\rangle-|-\sqrt{2}\alpha\rangle)/\sqrt{N_{\alpha}},
\end{equation}
by splitting it on a 50/50 beam splitter. Cat states of em fields have been produced inside micro-cavities \cite{haroche1,haroche2}, although the procedure used in these experiments works only for not too small values of $|\alpha|$.

Remarkably, for any value of $\alpha$ the state (\ref{start}) has one ebit of entanglement  \cite{hirota1,hirota2}. Indeed, as can be easily verified, the reduced density matrix ${\rm Tr}_1 |H_{\alpha}\rangle _{1,2}\langle H_{\alpha}|$ has two nonzero eigenvalues, both equal to 1/2.
Alternatively we may define
\begin{eqnarray}\label{+-}
|+\rangle&=&(|\alpha\rangle+|-\alpha\rangle)/\sqrt{N_+},\nonumber\\
|-\rangle&=&(|\alpha\rangle-|-\alpha\rangle)/\sqrt{N_-},\nonumber\\
N_{\pm}&=&2\pm 2c_{\alpha},
\end{eqnarray}
and note that
\begin{eqnarray}
|H_\alpha\rangle=(|+\rangle|-\rangle+|-\rangle|+\rangle)/\sqrt{2},
\end{eqnarray}
which manifestly has one ebit of entanglement.

For completeness we note one can construct a whole 
class of states with the same entanglement by applying local
unitary transformations to modes 1 and 2. Here we are interested only in transformations that take 
coherent states to coherent states. For instance,
the unitary displacement operator $D(\beta)$ acts on coherents states as follows \cite{qo},
\begin{equation}
D(\beta)|\alpha\rangle=\exp(i \phi_{\beta\alpha})|\alpha+\beta\rangle,
\end{equation}
with
\begin{equation}
\phi_{\beta\alpha}={\rm Im} (\beta\alpha^*).
\end{equation}
Furthermore, we can multiply the coherent state amplitude by an arbitrary phase factor. 
Thus, all states of the form
\begin{eqnarray}\label{family}
|H_{\alpha\beta\gamma\delta}\rangle\equiv
(|\alpha\rangle|\beta\rangle-\exp(i\Gamma_{\alpha\beta\gamma\delta})
|\gamma\rangle|\delta\rangle)/\sqrt{N_{\alpha_0}}
\end{eqnarray}
with
\begin{eqnarray}\label{family2}
|\alpha-\gamma|&=&|\beta-\delta|={\rm const}\equiv 2\alpha_0,\nonumber\\
\Gamma_{\alpha\beta\gamma\delta}&=&{\rm Im}(\beta\delta^*+\alpha\gamma^*),
\end{eqnarray}
can be produced from the state $|H_{\alpha_0}\rangle$  by the two operations mentioned above, 
and consequently possess one ebit of entanglement. 

Since the amount of entanglement in (\ref{start}) is independent of $\alpha$ it may seem that such states may be especially robust against decoherence due to photon absorption. Namely, although photon absorption attenuates the coherent state, that by itself has no effect on the entanglement. On the other hand, there will be decoherence due to the inevitable entanglement with the environment. That is, although a coherent state remains coherent, a superposition of coherent states does not remain coherent.  Precisely how much decoherence there is will be examined in the following Section.
A related issue, namely, using Schr\"odinger cat states to allow for correction of errors due to photon absorption, is discussed in \cite{cochrane}.
\section{Noise and decoherence}
We imagine a situation where a field mode 1 is ``stored'' in Alice's lab, while mode 2 describes a light beam propagating to Bob. Clearly, the latter mode will suffer from losses, but even Alice's system will not be lossless. For simplicity we then assume that both modes are equally lossy.  In particular,   
we assume the modes 1 and 2 travel each through a noisy channel characterized by
\begin{equation}\label{noise}
|\alpha\rangle_1|0\rangle_E\rightarrow |\sqrt{\eta}\alpha\rangle_1 
|\sqrt{1-\eta}\alpha\rangle_E,
\end{equation}
where the second state now refers to the ``environment'' and $\eta$ is the noise parameter, which gives the fraction of photons that survives the noisy channel. This noise model may look too simplistic but is in fact equivalent 
(after tracing over the environment) to a more complicated model where the environment is 
assumed to consist of many modes and where
\begin{equation}
|\alpha\rangle_1\Pi_i|0\rangle_{E_i}\rightarrow 
|\sqrt{\eta}\alpha\rangle_1  \Pi_{i}|\epsilon_i\alpha\rangle_{E_i},
\end{equation}
with $\sum_i |\epsilon_i|^2+\eta=1$. The model (\ref{noise}) describes realistic noise for optical fields. For micro-wave fields the influence of the thermal field may not be neglected, but for an optical field the temperature is effectively zero.

Starting from a state $|H_{\alpha\beta\gamma\delta}\rangle$ 
the density matrix of the modes 1 and 2 after traveling through the noisy channels
becomes
\begin{eqnarray}
\rho_{1,2}=
\frac{1}{N_{\alpha_0}}\left[|\sqrt{\eta}\alpha,\sqrt{\eta}\beta\rangle_{1,2}
\langle\sqrt{\eta}\alpha,\sqrt{\eta}\beta|
+|\sqrt{\eta}\gamma,\sqrt{\eta}\delta\rangle_{1,2}
\langle \sqrt{\eta}\gamma,\sqrt{\eta}\delta|\right.\nonumber\\\left.
-s|\sqrt{\eta}\alpha,\sqrt{\eta}\beta\rangle_{1,2}
\langle \sqrt{\eta}\gamma,\sqrt{\eta}\delta|
-s^*|\sqrt{\eta}\gamma,\sqrt{\eta}\delta\rangle_{1,2}
\langle \sqrt{\eta}\alpha,\sqrt{\eta}\beta|
\right].\end{eqnarray}
Here we abbreviated
\begin{equation}
|\alpha,\beta\rangle=|\alpha\rangle|\beta\rangle,
\end{equation}
and
\begin{equation}
s=\exp(-i\Gamma_{\alpha\beta\gamma\delta})\langle \sqrt{1-\eta}\gamma,\sqrt{1-\eta}\delta|
\sqrt{1-\eta}\alpha,\sqrt{1-\eta}\beta
\rangle.
\end{equation}
This decohered state obviously will have less entanglement than 1 ebit. A relevant measure of how much useful (to be used for teleportation for instance) entanglement is left, is  
the overlap of $\rho_{1,2}$ with states of the form (\ref{family}), since the latter always contain one ebit of entanglement.
The maximum overlap is in fact with the state $|H_{\sqrt{\eta}\alpha\sqrt{\eta}\beta\sqrt{\eta}\gamma\sqrt{\eta}\delta}\rangle$.
This defines a fidelity
\begin{eqnarray}
F&\equiv& \langle H_{\sqrt{\eta}\alpha\sqrt{\eta}\beta\sqrt{\eta}\gamma\sqrt{\eta}\delta} 
|\rho_{1,2} |H_{\sqrt{\eta}\alpha\sqrt{\eta}\beta\sqrt{\eta}\gamma\sqrt{\eta}\delta}\rangle
\nonumber\\
&=&\frac{[1-\exp(-4\eta\alpha_0^2)][1+\exp(-4(1-\eta)\alpha_0^2)]}
{2[1-\exp(-4\alpha_0^2)]},
\end{eqnarray}
with $\alpha_0$ as defined in (\ref{family2}).
Since this fidelity depends only on $\eta$ and $\alpha_0$ 
the whole family of states described by (\ref{family}) and
(\ref{family2})
decoheres in the same way. In Fig.~1 we plot the fidelity as a function of $\alpha_0$.
\begin{figure}\label{f1} \leavevmode
\epsfxsize=8cm \epsfbox{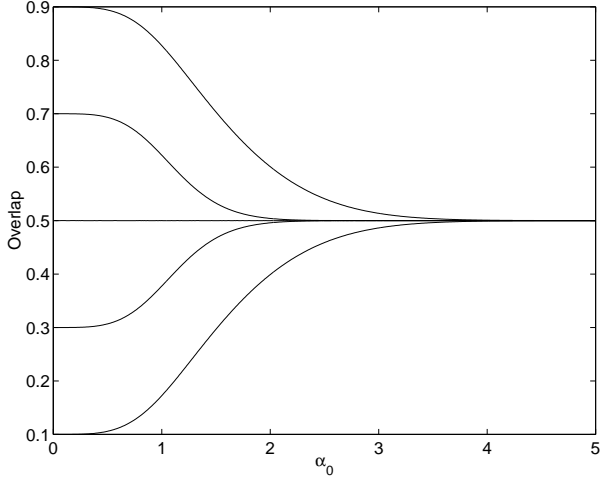} \caption{Overlap of $\rho_{1,2}$  with appropriately chosen fully entangled state as a function of $\alpha_0$ for various noise parameters $\eta=0.9,0.7,0.5,0.3,0.1$ for top to bottom curves.}
\end{figure}
From that figure one sees that 
for $\alpha_0\rightarrow 0$, $F\rightarrow\eta$.  For $\eta=1/2$, the fidelity is always $F=1/2$, and
the same fidelity is reached in the limit $|\alpha|\rightarrow\infty$ (for $\eta<1$).
Note that in this limit Alice and Bob actually obtain a separable state.

 For $\eta> 1/2$ the fidelity
decreases with increasing $\alpha_0$. For a fixed amount of noise $\eta>1/2$ the optimum states  
for teleportation are thus the family of states corresponding to $\alpha_0\rightarrow 0$, 
for which the state (\ref{start}) reduces to
\begin{equation}
|H_0\rangle\equiv (|0\rangle |1\rangle + |1\rangle |0\rangle)/\sqrt{2},
\end{equation}
in terms of photon number states. 
\section{Teleportation}
Any of the family of states (\ref{family}) can in principle be used to teleport one qubit. Indeed, as we will show now, Schr\"odinger cat
states of the form
\begin{equation}
|\psi_0\rangle=(\epsilon_+|\alpha\rangle+\epsilon_-|-\alpha\rangle)/\sqrt{N_0},
\end{equation}
with
\begin{equation}
N_0=|\epsilon_+|^2+|\epsilon_-|^2+2c_{\alpha}{\rm Re}\epsilon_-^*\epsilon_+,
\end{equation}
can be faithfully teleported using $|H_{\alpha}\rangle$.
The protocol we discuss here is chosen for its simplicity; it is not the optimal teleportation protocol. It does, however, lead with nonzero probability to perfect teleportation (if there is no noise). It is straightforward to formulate the optimum teleportation protocol using but it would require measurements that are hard to implement, see \cite{hirota2}.   
Here we let Alice mix her part of the entangled state (mode 1)
with the state to be teleported (in mode 0) by using a beam splitter.
The resulting state is\footnote{Phase factors $i$ have been eliminated, so that instead of the standard definition of the action of a beam splitter on coherent states:
\[ |\alpha\rangle|\beta\rangle\rightarrow |(\alpha+i\beta)/\sqrt{2}\rangle
|(i\alpha+\beta)/\sqrt{2}\rangle,\]
we get
\[ |\alpha\rangle|\beta\rangle\rightarrow |(\alpha+\beta)/\sqrt{2}\rangle
|(\alpha-\beta)/\sqrt{2}\rangle.\]
This corresponds to phase-shifting the second input and the second output modes by $-\pi/2$.} 
\begin{eqnarray}
|\psi_1\rangle_{0,1,2}&\equiv& \epsilon_+[|\sqrt{2}\alpha\rangle_0|0\rangle_1|\alpha\rangle_2
-[|0\rangle_0|\sqrt{2}\alpha\rangle_1|-\alpha\rangle_2]+\nonumber\\
&&-\epsilon_-[|-\sqrt{2}\alpha\rangle_0|0\rangle_1|-\alpha\rangle_2
-[|0\rangle_0|-\sqrt{2}\alpha\rangle_1|\alpha\rangle_2].
\end{eqnarray}
Subsequently Alice performs two photon number measurements on the modes 0 and 1 on her side. Denote the probability to find $n$ and $m$ photons in modes 0 and 1, respectively, by $P(n,m)$,
\begin{equation}\label{Pn}
P(n,m)=|_0\langle n|_1\langle m|\psi_1\rangle_{0,1,2}|^2.
\end{equation}
Only one of the two outcomes can be a nonzero number and let's
suppose that $n\neq 0$.
In this case the state on Bob's side collapses into
\begin{eqnarray}
|\psi'_0\rangle_2&=&(\epsilon_+|\alpha\rangle_2-(-1)^n\epsilon_-|-\alpha\rangle_2)/\sqrt{N'_0}\nonumber\\
N'_0&=&|\epsilon_+|^2+|\epsilon_-|^2-2(-1)^n c_{\alpha}{\rm Re}\epsilon_-^*\epsilon_+.
\end{eqnarray}
We see that provided $n$ is odd, the 
teleportation works perfectly. 
For $n$ even the transformation required for perfect state transfer
would be 
\begin{eqnarray}
|-\alpha\rangle&\rightarrow& -|-\alpha\rangle\nonumber\\
|\alpha\rangle&\rightarrow& |\alpha\rangle
\end{eqnarray}
which is not in general unitary (it is unitary only in the limit $|\alpha|\rightarrow\infty$).
Although for even nonzero values of $n$ the teleportation fidelity may still be better than can be achieved classically (i.e. without entanglement), we focus here only on the case $n$ is odd.
For $n$ odd (\ref{Pn})  reduces to
\begin{equation}
P(n,0)=\frac{|\langle n|\sqrt{2}\alpha\rangle|^2}{N_{\alpha}}\,\,{\rm for\, odd}\, n. 
\end{equation}
This quantity is independent of the state to be teleported, 
as is to be expected in a perfect teleportation protocol.
The probability of success (i.e. finding an odd number of photons in either mode) is
\begin{equation}
P_{{\rm odd}}=\sum _{n\,{\rm odd}} P(n,0)+\sum _{n\,{\rm odd}} P(0,n).
\end{equation}
These summations can be performed and the result turns out to be independent of $\alpha$,
\begin{equation}
P_{{\rm odd}}=\frac{1}{2}.
\end{equation}
\subsection{Noisy case}
When the noise parameter $\eta< 1$, the teleportation fidelity will drop below 1. On the other hand, as we will show, the probability to find an odd number of photons does not necessarily decrease below 50\%.

In the presence of noise, the states we should expect to be able to teleport more or less successfully are of the form
\begin{equation}\label{psi0}
|\tilde{\psi}_0\rangle=(\epsilon_+|\tilde{\alpha}\rangle+\epsilon_-|-\tilde{\alpha}\rangle)/\sqrt{\tilde{N}_0},
\end{equation}
with $\tilde{\alpha}=\sqrt{\eta}\alpha$ and
\begin{equation}
\tilde{N}_0=|\epsilon_+|^2+|\epsilon_-|^2+2c_{\tilde{\alpha}}{\rm Re} \epsilon_-^*\epsilon_+.
\end{equation}
Following the same steps as before, after Alice has measured $n\neq 0$ photons in mode 1, say, the joint entangled state of Bob's mode and the environment is
\begin{equation}
|\tilde{\psi'}_0\rangle=(\epsilon_+|\tilde{\alpha}\rangle|k\rangle-(-1)^n\epsilon_-|-\tilde{\alpha}\rangle|-k\rangle)
/\sqrt{N_k},
\end{equation}
where
\begin{eqnarray}
|k\rangle&\equiv& |\sqrt{1-\eta}\alpha\rangle|\sqrt{1-\eta}\alpha\rangle,\nonumber\\
N_k&=&|\epsilon_+|^2+|\epsilon_-|^2-(-1)^n2c_{\tilde{\alpha}}c^2_{\sqrt{1-\eta}\alpha} {\rm Re} 
\epsilon_-^*\epsilon_+.
\end{eqnarray}
We find again that a measurement of an odd number of photons is
required for near-perfect teleportation.
The probability of finding $n$ photons is now
\begin{equation}
P(n,0)=|\langle n|\sqrt{2}\tilde{\alpha}\rangle|^2\frac{N_k}{\tilde{N}_0 N_{\alpha}}
\,\,{\rm for\, odd}\, n,
\end{equation}
and the probability of finding an odd number of photons in either mode  is
\begin{equation}\label{P}
P_{{\rm odd}}=\sum_{n\,{\rm odd}} P(n,0)+\sum_{n\,{\rm odd}} P(0,n)=
\frac{1}{2}\frac{N_k N_{\tilde{\alpha}}}{\tilde{N}_0 N_{\alpha}}.
\end{equation}
Both these quantities
do depend now on the state to be
teleported through $N_k$ and $\tilde{N}_0$.
In principle there is, therefore, some information about the identity of the state $|\psi_0\rangle$ to be gained from the particular measurement outcome. This issue is further discussed in \cite{holger}. Here, however, we assume that Bob does not make any use of this extra knowledge about the identity of the state to be teleported. 
The fidelity of the teleported state is then simply given by
\begin{eqnarray}
F&\equiv& |\langle \tilde{\psi}_0|\tilde{\psi'}_0\rangle|^2\nonumber\\
&=&\frac{|A|^2+|B|^2+c_k(AB^*+BA^*)}{\tilde{N}_0N_k},
\end{eqnarray}
with 
\begin{eqnarray}
A&=&|\epsilon_+|^2+\epsilon_-^*\epsilon_+c_{\tilde{\alpha}},\nonumber\\
B&=&|\epsilon_-|^2+\epsilon_+^*\epsilon_-c_{\tilde{\alpha}}.
\end{eqnarray}
Suppose now we intend to teleport an arbitrary qubit in the space spanned by $|\alpha\rangle$ and $|-\alpha\rangle$. We may parametrize that qubit as follows,
\begin{eqnarray}\label{qubit}
|\psi_0(\theta,\phi)\rangle&=&\sin(\theta/2)|+\rangle+\cos(\theta/2)\exp(i\phi)|-\rangle,
\nonumber\\
0&\leq&\theta\leq \pi; \; 0\leq \phi<2\pi,
\end{eqnarray}
with the states $|\pm\rangle$ defined as in (\ref{+-}) and with a probability distribution ${\rm d}p=\sin\theta {\rm d}\theta {\rm d}\phi/(4\pi)$. This procedure corresponds to choosing a spin-1/2 representation for the qubit and choosing an arbitrary direction for its ``spin''.
The expectation value of the number of photons $N(\theta,\phi)$ in the state (\ref{qubit}) is
\begin{eqnarray}
N(\theta,\phi)
&=&\sin^2(\theta/2)|\alpha|^2\frac{N_-}{N_+}+\cos^2(\theta/2)|\alpha|^2\frac{N_+}{N_-}.
\end{eqnarray}
Averaging over $\theta$ and $\phi$ gives the average number of photons
\begin{eqnarray}
\langle N\rangle=\frac{1}{2}|\alpha|^2\left(\frac{N_-}{N_+}+\frac{N_+}{N_-}\right).
\end{eqnarray}
Although in the limit of small $\alpha$ the number of photons in both $|\alpha\rangle$ and $|-\alpha\rangle$ is small, a qubit (\ref{qubit}) has half a photon on average, since for $\alpha\rightarrow 0$, $\langle N\rangle\rightarrow 1/2$. For large $\alpha$ (roughly for $|\alpha|>1$) the average number of photons is simply $\langle N\rangle\approx|\alpha|^2$.

We can average the results for $P_{{\rm odd}}$  and the teleportation fidelity $F$ over $\theta$ and $\phi$. Averaging (\ref{P}) gives, surprisingly perhaps,
\begin{equation}
\langle P_{{\rm odd}}\rangle=\frac{1}{2},
\end{equation}
just as in the noiseless case. The average teleportation fidelity is plotted in Fig.~2.
\begin{figure}\label{f2} \leavevmode
\epsfxsize=8cm \epsfbox{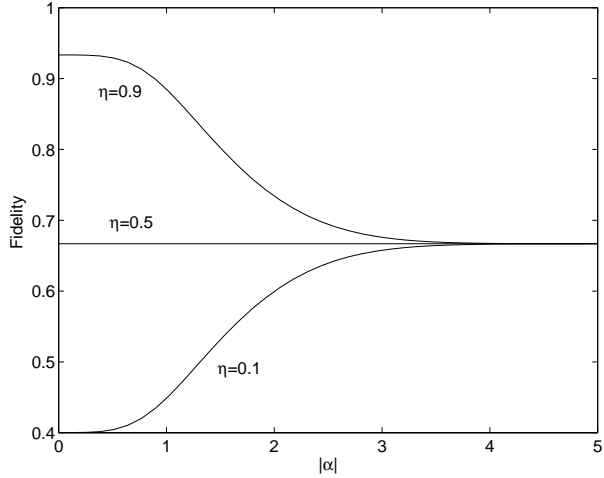} \caption{Teleportation fidelity in the noisy case as a function of $|\alpha|$ for different values of the noise parameter $\eta$.}
\end{figure}
For $|\alpha|\gg 1$ the average teleportation fidelity approaches 2/3 for any noise parameter $\eta<1$, which is the value expected for ``classical'' teleportation, where Alice performes a projective measurement in the basis $|+\rangle,|-\rangle$, tells Bob the result, who then prepares that state. 
This is because in this limit Alice and Bob share no entanglement.
\subsection{Using different entangled states}
States of the form
\begin{equation}\label{alt}
|G_{\alpha}\rangle_{1,2}=(|\alpha\rangle_1|\alpha\rangle_2+|-\alpha\rangle_1|-\alpha\rangle_2),
\end{equation}
as discussed in \cite{yeazell,filip}
have less than one ebit of entanglement \cite{hirota2}. In fact, the reduced density matrix has two nonzero eigenvalues given by
\begin{equation}
\lambda_{\pm}=\frac{(1\pm c_{\alpha})^2}{2+2c_{\alpha}^2}
\end{equation}
and its entanglement is $E=-\sum_{i=\pm}\lambda_i\log_2\lambda_i\leq 1$, with the equality sign holding only in the limit $|\alpha|\rightarrow\infty$.
We therefore expect this type of state to be less successful in teleporting quantum states. Indeed, using the same teleportation procedure as before, we find that the probability to find a desired measurement outcome is less than 1/2 now, although the teleportation fidelity still is $100\%$ in the absence of noise.
In particular, following the same steps as before, after Alice measures a nonzero number of photons in mode 1, Bob's state collapses into
\begin{eqnarray}
|\psi'_0\rangle_2&=&(\epsilon_+|\alpha\rangle_2+(-1)^n\epsilon_-|-\alpha\rangle_2)/\sqrt{N'_0}\nonumber\\
N'_0&=&|\epsilon_+|^2+|\epsilon_-|^2+2(-1)^n c_{\alpha}{\rm Re}\epsilon_-^*\epsilon_+.
\end{eqnarray}
Hence, a successful teleportation now occurs when Alice measures a nonzero {\em even} number of photons. The probability of succesful teleportation is
\begin{equation}
P_{{\rm even}}=\sum _{n>0\,{\rm even}} P(n,0)+\sum _{n>0\,{\rm even}} P(0,n)=
\frac{(1-c_\alpha)^2}{2+2c_\alpha^2},
\end{equation}
which is less than 1/2. 
For small $|\alpha|$ this probability approaches zero: there is indeed no entanglement in the state $|G_{\alpha}\rangle$ for $|\alpha|\rightarrow 0$. For $|\alpha|\rightarrow\infty$, on the other hand, the probability of successful teleportation becomes 1/2 as before, and correspondingly the state
$|G_{\alpha}\rangle$ does indeed possess one ebit of entanglement in that limit.
In Fig.~3 we plot both the entanglement and $P_{{\rm even}}$ as functions of $|\alpha|$.
\begin{figure}\label{f3} \leavevmode
\epsfxsize=8cm \epsfbox{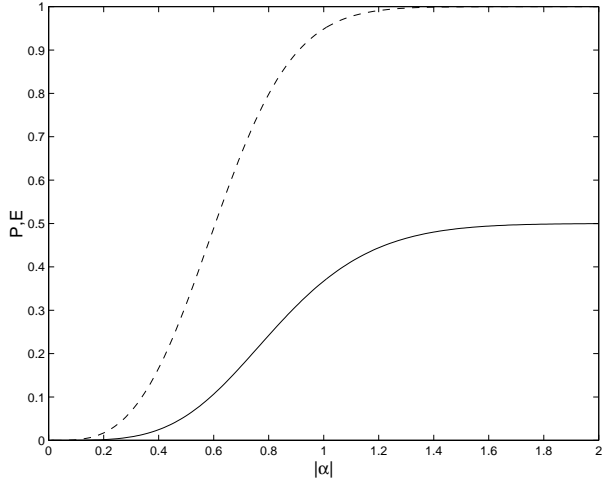} \caption{Probability of successful teleportation $P_{{\rm even}}$ (solid curve)
with a state of the form (\ref{alt}) and its entanglement $E$ (dashed curve) as functions of $|\alpha|$ in the absence of noise.}
\end{figure}
\section{Conclusions}
We studied properties of entangled coherent states of the form
\[|\alpha\rangle|\alpha\rangle-|-\alpha\rangle|-\alpha\rangle.\]
These states can be produced from Schr\"odinger cat states by using a 50/50 beam splitter.
Such states possess exactly one ebit of entanglement\cite{hirota1} and can be used to teleport one qubit encoded in superpositions of $|\alpha\rangle$ and $|-\alpha\rangle$. We discussed a simple protocol that achieves this aim with a 50\% probability of success. This protocol requires beam splitters, the ability to produce Schr\"odinger cat states, and photon counting.
A teleportation experiment along the lines sketched here may be feasible in the context of cavity-QED.
In the presence of noise, the teleportation fidelity decreases, but the probability of success does not necessarily decrease and its average value in fact remains constant. 
\section*{Acknowledgements}
We thank Chris Fuchs and Masahide Sasaki for useful discussions and comments.


\begin{thebibliography}{99}

\bibitem{qkd}A. Ekert, Phys. Rev. Lett. {\bf 67}, 661 (1991).

\bibitem{telep} C.H. Bennett, H.G. Brassard, C. Crepeau,
R. Jozsa, A. Peres, and W.K. Wootters,
Phys. Rev. Lett., {\bf 70}, 1895 (1993).

\bibitem{dik}
D. Bouwmeester, J.W. Pan, K. Mattle, M. Eibl, H. Weinfurter, and A. Zeilinger,
Nature {\bf 390}, 575 (1997).

\bibitem{ent}C.H. Bennett, D.P. DiVincenzo, J.A. Smolin, and W.K. Wootters, Phys. Rev. A {\bf 54}, 3824 (1996).

\bibitem{akira}A. Furusawa, J.L. S\o rensen, S.L. Braunstein, C.A. Fuchs, H.J. Kimble, and E.S. Polzik, Science {\bf 282}, 706 (1998).

\bibitem{qo}D.F. Walls and G.J. Milburn, {\em Quantum Optics} (Springer-Verlag, Berlin, 1994).

\bibitem{enk}S.J. van Enk, Phys. Rev. A {\bf 60}, 5095 (1999). 

\bibitem{hirota1} O. Hirota and M. Sasaki,
Proceedings QCMC, 2000, to be published.

\bibitem{hirota2} O. Hirota, K. Nakamura, M. Sohma, and K. Kato,
to be published.

\bibitem{chai}C.L. Chai, Phys. Rev. A {\bf 46}, 7187 (1992).

\bibitem{knight}The nontrivial character of the relative phase $\phi$ in the entangled state
\[|\alpha\rangle|\alpha\rangle+\exp(i\phi)|-\alpha\rangle|-\alpha\rangle\]
can be seen already by considering superpositions of coherent states
\[|\sqrt{2}\alpha\rangle+\exp(i\phi)|-\sqrt{2}\alpha\rangle,\]
from which the entangled states can be produced by using a 50/50 beamsplitter.
For instance, for $\phi=0$ this state is a so-called even coherent state, containing superpositions of states with even numbers of photons. Similarly, for $\phi=\pi$ we get the odd coherent state, with always an odd number of photons. For $\phi=\pi/2$ on the other hand, the photon number distribution is Poissonian, just as for a coherent state. For a review of nonclassical properties of these types of states, see V. Bu\v{z}ek and P.L. Knight, Prog. in Optics {\bf XXXIV}, p. 1, Elsevier Amsterdam (1995). 

\bibitem{yeazell}J.C. Howell and J.A. Yeazell, Phys. Rev. A {\bf 62}, 012102 (2000).

\bibitem{filip}R. Filip, J. \v{R}eha\v{c}ek and M. Du\v{s}ek, quant-ph/0011006.

\bibitem{sanders}B.C. Sanders, Phys. Rev. A {\bf 45}, 6811 (1992).

\bibitem{munro}W.J. Munro, G.J. Milburn and B.C. Sanders, Phys. Rev. A {\bf 62}, 052108 (2000).


\bibitem{haroche1} M. Brune, E. Hagley, J. Dreyer, X. Maitre, C. Wunderlich, J.M. Raimond, and S. Haroche, Phys. Rev. Lett. {\bf 77}, 4887 (1996).

\bibitem{haroche2} J.M. Raimond, M. Brune, and S. Haroche, Phys. Rev. Lett. {\bf 79}, 1964 (1997).

\bibitem{cochrane}P.T. Cochrane, G.J. Milburn and W.J. Munro, Phys. Rev. A {\bf 59}, 2631 (1999).

\bibitem{holger}
 H.F. Hofmann, T. Ide, T. Kobayashi, and A. Furusawa, Phys. Rev. A {\bf 62}, 062304 (2000)

\end{thebibliography}
\end{document}